\documentclass[aps,prl,letterpaper,amsmath,amssymb,superscriptaddress,nofootinbib,twocolumn,floatfix]{revtex4-1}

\usepackage{graphicx}
\usepackage[usenames]{color}
\usepackage[colorlinks=true,linkcolor=blue,citecolor=blue,anchorcolor=green,urlcolor=blue]{hyperref}

\usepackage{ulem}

\usepackage[percent]{overpic}
\usepackage{tabularx}

\usepackage{pdfpages}
\usepackage{pgffor}
\makeatletter
\AtBeginDocument{\let\LS@rot\@undefined}
\makeatother

\def\be{\begin{equation}}
\def\ee{\end{equation}}
\def\bea{\begin{eqnarray}}
\def\eea{\end{eqnarray}}

\begin{document}

\title{
Confinement of many-body Bethe strings
}

\author{Jiahao Yang}
\affiliation{Tsung-Dao Lee Institute, Shanghai Jiao Tong University, Shanghai 201210, China}

\author{Tao Xie}
\affiliation{Neutron Scattering Division, Oak Ridge National Laboratory, Oak Ridge, TN 37831, USA}

\author{S. E. Nikitin}
\email{stanislav.nikitin@psi.ch}
\affiliation{Quantum Criticality and Dynamics Group, Paul Scherrer Institut, CH-5232 Villigen-PSI, Switzerland}

\author{Jianda Wu}
\email{wujd@sjtu.edu.cn}
\affiliation{Tsung-Dao Lee Institute, Shanghai Jiao Tong University, Shanghai 201210, China}
\affiliation{School of Physics and Astronomy, Shanghai Jiao Tong University, Shanghai 200240, China}
\affiliation{Shanghai Branch, Hefei National Laboratory, Shanghai 201315, China}

\author{A. Podlesnyak}
\affiliation{Neutron Scattering Division, Oak Ridge National Laboratory, Oak Ridge, TN 37831, USA}

\begin{abstract}
Based on Bethe-ansatz approach and inelastic neutron scattering experiments,
we reveal evolution of confinement of many-body Bethe strings in ordered regions of
quasi-one-dimensional antiferromagnet $\rm YbAlO_3$.
In the antiferromagnetic phase, the spin dynamics is dominated
by the confined length-1 Bethe strings, whose dominancy
in the high-energy branch of the excitation spectrum
yields to the confined length-2 Bethe strings when the material is tuned
to the spin-density-wave phase.
In the thermal-induced disordered region,
the confinement effect disappears,
and the system restores the conventional quantum
integrable physics of the one-dimensional Heisenberg model.
Our results establish a unified picture based on Bethe string for the spin dynamics in different magnetic phases of $\rm YbAlO_3$,
and thus provide profound insight into the many-body quantum magnetism.
\end{abstract}
\date{\today}
\maketitle

{\it Introduction.~}The one-dimensional (1D) spin-1/2 Heisenberg model, a paradigmatic model
for studying quantum many-body physics, exhibits rich magnetic excitations
such as magnon \cite{bloch_zur_1930,holstein_field_1940,karabach_introduction_1997}, spinon \cite{Jimbo1995, caux_two-spinon_2008,castillo_exact_2020,caux_four-spinon_2006,mourigal_fractional_2013}, (anti)psinon \cite{karbach_introduction_2000},
and Bethe strings
\cite{bethe1931,takahashi_1D_1971,Gaudin_XXZ_1971,Taka_suzuki_XXZ_1972,Takahashi1999,yang_string_1D_2019,wang_string_experimental_2018,Wang_quantum_2019,bera_string_dispersions_2020,kohno_string_dynamically_2009}.
Though the former three types of excitations have been well
studied via both theory and experiments,
the Bethe strings, exotic many-magnon bound states,
are long-sought in real materials.
Recently,
with the aid of Bethe-ansatz calculation
\cite{yang_string_1D_2019},
substantial progress has been made by THz spectroscopy and inelastic
neutron scattering (INS) experiments on the
quasi-1D Heisenberg-Ising antiferromagnet $\rm SrCo_2V_2O_8$ (SCVO)
\cite{wang_string_experimental_2018,bera_string_dispersions_2020} and $\rm BaCo_2V_2O_8$ (BCVO)
\cite{Wang_quantum_2019}.
The progress reveals the existence of the Bethe strings in SCVO and BCVO and identifies their vital contributions to spin dynamics.

\begin{figure}[t]
    \includegraphics[width=0.8\columnwidth]{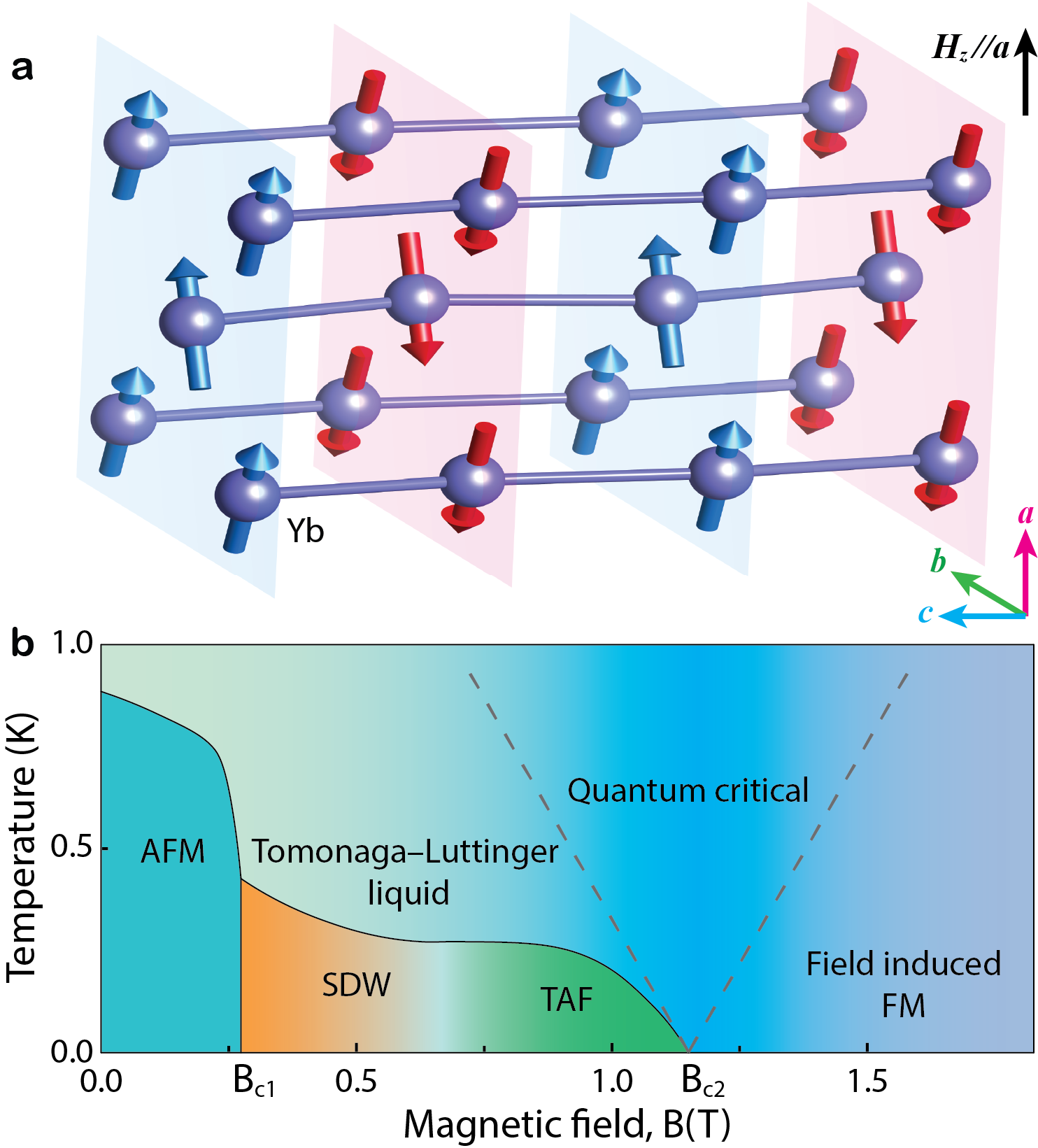}
    \caption{Schematic illustration of the magnetic structure and phase diagram of $\rm YbAlO_3$.
    \textbf{a}
    Below N\'eel temperature $T_N$, the magnetic moments of Yb ions are antiferromagnetically (AFM) ordered along chain direction (the $c$-axis),
    and the interchain coupling is ferromagnetic (FM) in the ab-plane.
    \textbf{b} Field-temperature phase diagram with AFM phase, spin-density-wave (SDW) phase, and presumably transverse-antiferromagnetic (TAF) phase, indicated by blue, orange, and green, respectively.
    Two gray dotted lines show the crossover regions between Tomonaga-Luttinger liquid, quantum critical, and field-induced FM.
    }
    \label{fig:Struc_phase_diag}
\end{figure}

Recently, a rare-earth-based material, $\rm YbAlO_3$ (YAO), is evidenced to be a quasi-1D Heisenberg antiferromagnet
\cite{wu_TLL_2019,podlesnyak_lowenergy_2021}.
The magnetic ion $\rm Yb^{3+}$ carries on an effective spin $S=1/2$ due to the combined effect between the spin-orbit coupling and the crystal field effect.
And among $\rm Yb^{3+}$ ions, the dominant spin exchange interaction ($J$) is isotropic along the chain direction (the $c$ direction in Fig.~\ref{fig:Struc_phase_diag}{\bf a})
\cite{wu_TLL_2019,nikitin_multiple_2021}.
The Tomonaga-Luttinger liquid (TLL) theory has been successfully
applied to understand ground state and
low-energy excitations of the material
but failed to explain
the physics for the observed spectrum with rich structure
beyond low-energy region
\cite{wu_TLL_2019,nikitin_multiple_2021}.
Thus a unified physical picture beyond TLL is desired to give a complete understanding for the complete spectral response.

In this letter by the combined efforts of the INS experiment and the Bethe-ansatz calculations,
we report that the Bethe strings can provide a unified physical picture to give a full understanding
for all spin dynamical spectra observed in different regions of YAO.
Specifically, in ordered regions, the confined length-1 strings
dominate the dynamic spectrum in the antiferromagnetic (AFM) phase, which gives way to the confined
length-2 strings for the high energy branch of the spin dynamic when
the material turns into the spin-density-wave (SDW) phase.
In the thermal-induced disordered region Bethe strings are released from confinement.
Our work not only uncovers the existence and confinement of Bethe strings
but also provides a unified picture based on Bethe string 
to quantitatively describe the spin dynamics in different phases of YAO,
which is beyond the conventional static order-parameter paradigm \cite{Fan_Phase_2020} and low-energy effective theories \cite{wu_TLL_2019,nikitin_multiple_2021,tomonaga_remarks_1950,Luttinger_1960,Luther_TLL_1975,Starykh_q1DSDW_2014,Essler_q1DHeisenberg_1997}.

{\it Experimental details.~}The inelastic neutron scattering (INS) experiments were performed at the time-of-flight Cold Neutron Chopper Spectrometer (CNCS)~\cite{CNCS1,CNCS2} at the Spallation Neutron Source, Oak Ridge National Laboratory. A YAO crystal used in previous studies~\cite{wu_TLL_2019, nikitin_multiple_2021} with a mass of $\sim$~0.5~g was mounted in the $(0~K~L)$ scattering plane. The zero field data were collected in a standard cryostat equipped with a dilution insert. The 0.4~T dataset was measured in a vertical 5~T cryomagnet with a dilution refrigerator and field was applied along the [1 0 0] direction (the easy axis). To obtain optimal coverage in the energy and momentum space, the sample was rotated along the vertical axis by 90$^{\circ}$. The data were collected with an incident neutron energy $E_{\mathrm{i}}$ = 1.55 meV, with an energy resolution of 0.05~meV at the elastic line. We used the software packages~\textsc{MantidPlot}~\cite{Mantid} and ~\textsc{Horace}~\cite{Horace} for data reduction and analysis. The actual temperature of the sample was calculated using the detailed balance principle, $S(\mathbf{Q}, E) = e^{-E/k_{\rm B}T} S(\mathbf{Q}, -E)$, by comparing the spectral intensity at positive and negative energy transfer.

{\it Theoretical Models.~}The phase diagram of YAO (Fig.~\ref{fig:Struc_phase_diag}{\bf b}) is obtained
from specific heat and magnetization measurements \cite{nikitin_multiple_2021,wu_TLL_2019,Fan_Phase_2020},
which is qualitatively similar to that of SCVO \cite{bera_string_dispersions_2020} and BCVO \cite{Giant_2015}.
At low temperatures, the interchain coupling helps to stabilize
different long-range orders (LROs) in the material.
By tuning the external magnetic field,
the material can change from the AFM phase to
the SDW phase after crossing through the critical field $B_{c1} = 0.32$~T.
With the field further increasing to about 0.7 T the system begins to transform from the SDW phase to the presumably
transverse-antiferromagnetic (TAF) phase and eventually is fully polarized when $B>B_{c2} = 1.15$~T.
In the disordered region of YAO, the interchain coupling can be neglected following the chain mean-field treatment,
and the dominant magnetic property in YAO can then be described by the spin-1/2 Heisenberg model
with external longitudinal field $H_z$
(along $a$ direction in
Fig.~\ref{fig:Struc_phase_diag}{\bf a}),
\be
H_0 = J \sum_i \mathbf{S}_i\cdot\mathbf{S}_{i+1}
    - H_z S_i^z,
\label{eq:H0_XXX_h}
\ee
where the AFM coupling $J=0.21$~meV, and $\mathbf{S}_i$ is the spin operator
at site $i$ with spin components $S^{\alpha}_i\ (\alpha=x,y,z)$.
In the ordered phases, different LROs are characterized by different
ordering wavevector $Q=(1-m)\pi$ where magnetization density $m$ is
the ratio of magnetization $M_z$ to its saturation value $M_s$.
The LRO can effectively induce a mean field ${\bf{h}}_{z} = h_Q \sum_i \cos(Q r_i)\hat{z}$ which couples to
the spin chain Eq.~(\ref{eq:H0_XXX_h})
\cite{wu_TLL_2019,nikitin_multiple_2021,Essler_q1DHeisenberg_1997,Starykh_q1DSDW_2014},
\be
H=H_0 - h_Q \sum_i \cos(Q r_i) S^z_i,
\label{eq:H_hic}
\ee
where field strength $h_Q$ is generally dependent on
the strength of LRO.
Apart from the AFM phase, the effective mean field ${\bf{h}}_{z}$
in general is incommensurate,
and becomes commensurate when $2\pi/Q$ is a rational number
(for example, ${\bf{h}}_{z}$ is commensurate in AFM phase with $Q = \pi$).
In this letter, following
Bethe ansatz approach we analytically study the spin dynamics based on the
effective Hamiltonians of Eq.~(\ref{eq:H0_XXX_h}) and
Eq.~(\ref{eq:H_hic}) for disordered and ordered regions,
respectively. The obtained results are then
compared in detail with the INS results, which reveals
rich spin dynamics as discussed below.
Note, due to the strong Ising-anisotropy of $g$ tensor in YAO the main contribution to the INS spectrum comes from the longitudinal spin component~\cite{wu_TLL_2019}, which makes interpretation of the spectral function much easier compared with BCVO and SCVO with more isotropic $g$ factor of the ground state doublet.

\begin{figure}[t]
    \includegraphics[width=\columnwidth]{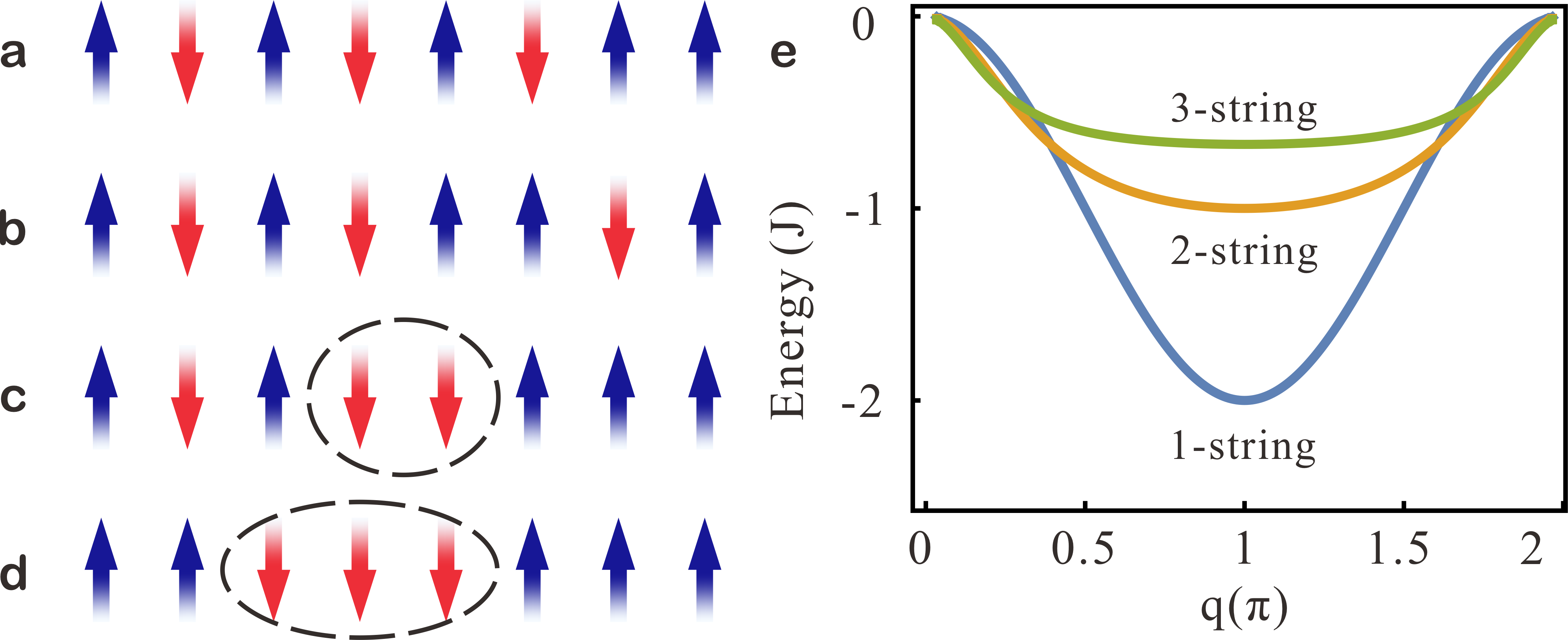}
    \caption{Pictorial spin configurations for different quasiparticles of 1D spin-1/2 Heisenberg model and their band structures.
    \textbf{a}, Ground state (GS) with SDW ordering and finite magnetization $M_z = N/2 - M$, where $M$ is the number of down spins with respect to the fully up-polarized state.
    \textbf{b-d}, Excitations generated from the GS. Psinon-antipsinon pair $\psi\psi^*$ (\textbf{b}), two-string $\chi^2$ (\textbf{c}), and three string $\chi^3$ (\textbf{d}).
    Viewing the down spin (red arrow) and up spin (blue arrow) as the magnon and vacuum respectively, many-body Bethe string $\chi^j$ ($j\geq2$) contains $j$ bounded magnons and lives in the corresponding $j$-string band in \textbf{e}.
    The 1-string band accommodates unbound magnons, and GS has the lowest total energy in the 1-string band.
    }
    \label{fig:spin_config}
\end{figure}

\begin{figure*}[t]
    \centering
    \includegraphics[width=2\columnwidth]{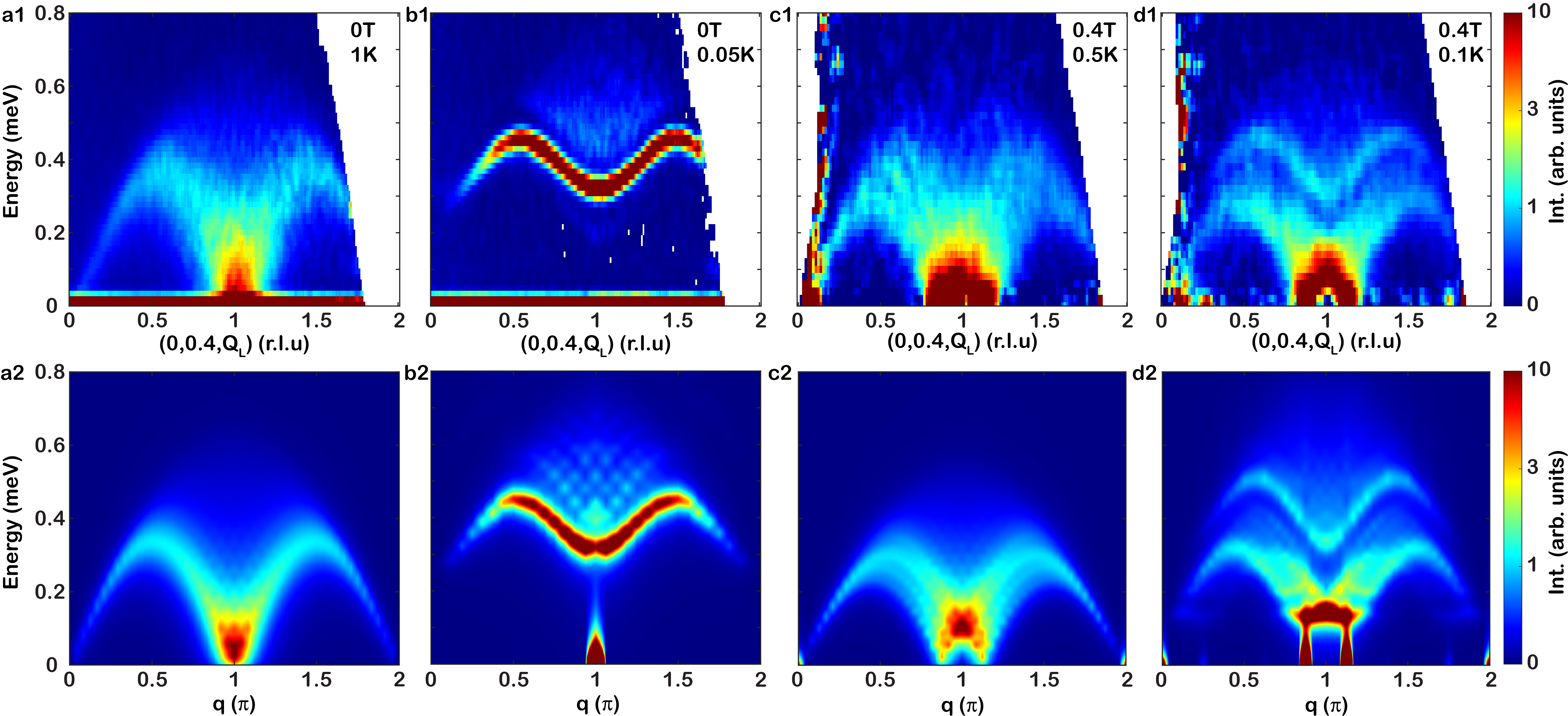}
    \caption{
    Spin dynamics for YAO.
    INS spectra without magnetic fields were collected at 1 K (\textbf{a1}) and 0.05 K (\textbf{b1}).
    The data in (\textbf{a1}) and (\textbf{b1}) were integrated within $K = [0.3,0.5]$, $H = [-0.1,0.1]$ r.l.u..
    No background subtraction was applied to the data.
    The INS spectra at $B=0.4$~T at 0.5 K (\textbf{c1}) and 0.1 K (\textbf{d1}).
    The data were integrated within all available $\mathbf{Q}$ ranges along the orthogonal directions and the background measured at 3~T was subtracted from both spectra.
    Correspondingly, thermal DSFs of $H_0$ are determined in \textbf{a2} and \textbf{c2},
    and zero temperature DSFs of $H$ are determined in \textbf{b2} and \textbf{d2}
    where the ordering fields are $h_\pi=0.27J$ and $h_Q=0.35J$ respectively.
    }
    \label{fig:DSF_parts}
\end{figure*}

{\it Dynamical structure factor.~}Following standard linear response theory
\cite{Negele1988,chaikin_lubensky_1995,Zhu_MTCMM_2005},
the thermal dynamical structure factor (DSF) for spin
along longitudinal ($z$) direction is given by
($k_B=1$)
\be
S^{zz}(T,q,\omega)=\frac{2\pi}{\mathcal{Z}}
\sum_{\lambda,\mu}
e^{- E_\lambda/T }
|\langle \lambda|S_q^z|\mu \rangle|^2
\delta(\hbar \omega-E_\mu+E_\lambda),
\label{eq:DSF_T}
\ee
with partition function $\mathcal{Z}$,
transfer momentum $q$ and transfer energy $\hbar\omega$ between two eigenstates $|\lambda\rangle$ and $|\mu\rangle$ with eigenenergies $E_\lambda$ and $E_\mu$.
The double summation goes over all the eigenstates of the effective Hamiltonians of
Eq.~\eqref{eq:H0_XXX_h} or Eq.~\eqref{eq:H_hic}.
At zero temperature the system is in the ground state,
i.e. $|\lambda\rangle=|GS\rangle$,
and the thermal DSF reduces to a single summation,
\be
S^{zz}(q,\omega)=2\pi
\sum_{\mu}
|\langle GS|S_q^z|\mu \rangle|^2
\delta(\hbar\omega-E_\mu+E_{GS}).
\label{eq:DSF_no_T}
\ee
In the following, we focus on the transfer energy range
$\hbar\omega \leq 0.8$ meV in compliance with the experimental data.

{\it Analysis and Discussions.~}For the effective Hamiltonian $H_0$ of YAO,
its excitations can be exactly obtained from the Bethe-ansatz method  \cite{franchini_introduction_2017}.
In general, the excitation can be decomposed into
Bethe strings with different lengths.
Typically, a Bethe string $\chi^{j}$ ($j\geq2$) of length $j$ contains $j$ bounded magnons,
referred to as $j$-string.
While if $j=1$, the 1-string $\chi^{1}$ is just unbound magnon, as shown in Fig.~\ref{fig:spin_config}.
The psinon ($\psi$) and antipsinon ($\psi^*$) can be understood as the ``particle'' and ``hole'' excitations from the ground state in the 1-string band,
which are always created in pairs ($\psi\psi^*s$).
And the psinon-antipsinon (PAP) pairs can adiabatically connect to fractionalized fermionic spinons at $m = 0$ and bosonic magnons at $m \simeq1$
\cite{karbach_introduction_2000}.
All the excitations can be
regarded as the combination of these quasiparticles,
i.e. $n\chi^j n'\psi \psi^*$
with integers $n,\ n' \geq0$.
The obtained dynamical spectra can produce characteristic spectrum continua due to the many-body nature of Bethe states.
Although the effective Hamiltonian $H$ is different from $H_0$ by an effective field $\textbf{h}_z$,
it is beyond the integrability and can not be solved exactly.
Here we tackle this problem by the truncated string state space method \cite{Supp_Mat}.
Following this method the thermal DSF is calculated to
describe the data collected at 1 K and 0.5 K,
while zero-temperature DSFs are obtained to compare with
data at 0.05 K and 0.1 K.

In the disordered region with $B=0$~T (corresponding to $m=0$),
the dominant low-energy excitations are
1-strings.
When sample temperature $T_{\rm s}=1\ \text{K}\sim0.4J$, the thermal fluctuations are non-negligible
and needs to be considered in the calculation.
From thermal DSF Eq.~\eqref{eq:DSF_T}
it is straightforward to observe that among thermal sampling states
$|\lambda\rangle$
the spectral weight
is exponentially suppressed with increasing eigenenergy $E_\lambda$
due to the Boltzmann factor.
As such it is reasonable to select a cutoff $E_\lambda \leq 4T_{\rm s}$
for thermal sampling
states $|\lambda\rangle$ (i.e. Boltzmann factor $\lesssim 1.8\%$).
With $T_{\rm s}=1$ K and $m=0$,
the thermal DSF Eq.~\eqref{eq:DSF_T} of $H_0$
shows that 1-strings dominate the dynamical spectrum (Fig.~\ref{fig:DSF_parts}{\bf a2})
and exhibits a large thermal broadening effect
compared with zero temperature result 
(see Supplemental Material \cite{Supp_Mat}),
and the INS spectrum
(Fig.~\ref{fig:DSF_parts}{\bf a1})
agrees well with the theoretical spectrum (Fig.~\ref{fig:DSF_parts}{\bf a2}).
In the AFM phase,
the zero temperature DSF Eq.~\eqref{eq:DSF_no_T} of $H$ with $h_\pi=0.27J$ shows that
1-strings are confined by the effective staggered field
and the corresponding INS spectrum  (Fig.~\ref{fig:DSF_parts}{\bf b1}) is confirmed by the theoretical result
(Fig.~\ref{fig:DSF_parts}{\bf b2}).
Note that the Bragg peak at zero energy is present in our neutron scattering data at $\mathbf{Q} = (0~0~1)$, but not visible in Fig.~\ref{fig:DSF_parts}{\bf b2} because of the selected integration range, $K = [0.3, 0.5]$ r.l.u.
It is worth noting that spinon confinement observed in
SCVO \cite{Wang_confined_2016,Bera_spinon_confine_2017},
BCVO \cite{faure_topological_2018}
and $\rm CaCu_2O_3$ \cite{lake_confinement_2010} is also the confinement of 1-strings.
However, the spinon picture only works at $m=0$ and is invalid when $m>0$ \cite{Jimbo1995,caux_two-spinon_2008,castillo_exact_2020,caux_four-spinon_2006}, in contrast, the Bethe string picture can describe excitations for both of $m=0$ and $m>0$.
Thus, we shall continuously discuss our results based on Bethe strings in the following.

In the disordered phase with $B=0.4$~T (corresponding to $m\simeq12\%$),
2-strings $\chi^2$ are also favored in the spin dynamics
in addition to 1-strings \cite{Supp_Mat,kohno_string_dynamically_2009}.
When $T_{\rm s}=0.5\ \text{K}\sim0.2J$,
the thermal fluctuation can not be neglected.
Following the same strategy in the case of $T_{\rm s}=1$ K,
thermal DSF Eq.~\eqref{eq:DSF_T} is determined for $H_0$ at $T_{\rm s}$ and $m\simeq12\%$ with energy cutoff $E_\lambda\leq4T_{\rm s}$.
In addition, with the presence of external field,
the dipole interaction
in $ab$ plane would deviate from the ``magic'' line of
vanishing dipole interaction \cite{WLS_AF_2019},
leading to non-negligible 3D fluctuations
that further frustrate the intrachain couplings
within the framework of the chain mean-field theory.
From theoretical calculations, we find that
at $B=0.4$ T
the obtained DSF with adjusted $J=0.18$ meV
(Fig.~\ref{fig:DSF_parts}{\bf c2})
is consistent
with the INS experimental spectrum (Fig.~\ref{fig:DSF_parts}{\bf c1}).
We note that the continuum of 2-strings is masked by that of 1-strings
\cite{Supp_Mat}
and there is no clear energy gap between these two continua
in Fig.~\ref{fig:DSF_parts}\textbf{c1,c2}.
In fact, this energy gap can be tuned by external magnetic field $H_z$, Ising-anisotropy of the Heisenberg-Ising model, and
next-nearest-neighbor coupling $J_2$ in various DSFs.
\cite{kohno_string_dynamically_2009,yang_string_1D_2019,Anna_dynamical_2020}.

In the SDW phase with $B=0.4$ T,
the observed INS spectrum (Fig.~\ref{fig:DSF_parts}{\bf d1})
reveals rich structures of string excitations from low to high energy.
With $h_Q=0.35J$, $J=0.18$ meV and $m\simeq12\%$, 
the zero temperature DSF Eq.~\eqref{eq:DSF_no_T}
of $H$ shows that both 1- and 2-strings are confined by the
effective field ${\bf h}_z$,
which plays the same role as that of the staggered field
for the confinement in the AFM phase.
Accordingly,
confined 2-strings exhibit a characteristic M-shaped high-energy continuum
that is gapped from the low-energy counterpart of confined 1-strings (Fig.~\ref{fig:DSF_parts}{\bf d2}) \cite{Supp_Mat}.
The appearance of a ``gap'' between these two continua reflects
the energy cost for the confinement of the 2-strings compared with the deconfined case
(Fig.~\ref{fig:DSF_parts}{\bf c1,c2}).
Both the characteristic gap and the shape of continua can be quantitatively
compared between
experimental (Fig.~\ref{fig:DSF_parts}{\bf d1})
and
theoretical results (Fig.~\ref{fig:DSF_parts}{\bf d2}),
which explicitly confirms the confinement of the Bethe strings of spin chains in YAO.
In addition,
at zero transfer energy,
a series of satellite peaks appear at $n Q\ (n=1,2,\cdots)$
in Fig.~\ref{fig:DSF_parts}{\bf d2}
\cite{Supp_Mat}.
This is because
the effective field ${\bf h}_z$ can connect two Bethe states with momentum difference $\Delta q= Q$
and stabilizes an SDW-ordered ground state.
And this ground state has all
the Bethe states with momentum being integer times of $Q$,
which makes possible transitions at many different momentum-transfer.
Therefore, the satellite peaks only occur at $q=nQ$, which is consistent with the multi-fermion scattering mechanism \cite{nikitin_multiple_2021}.

\begin{figure}[t]
    \centering
    \includegraphics[width=0.9\columnwidth]{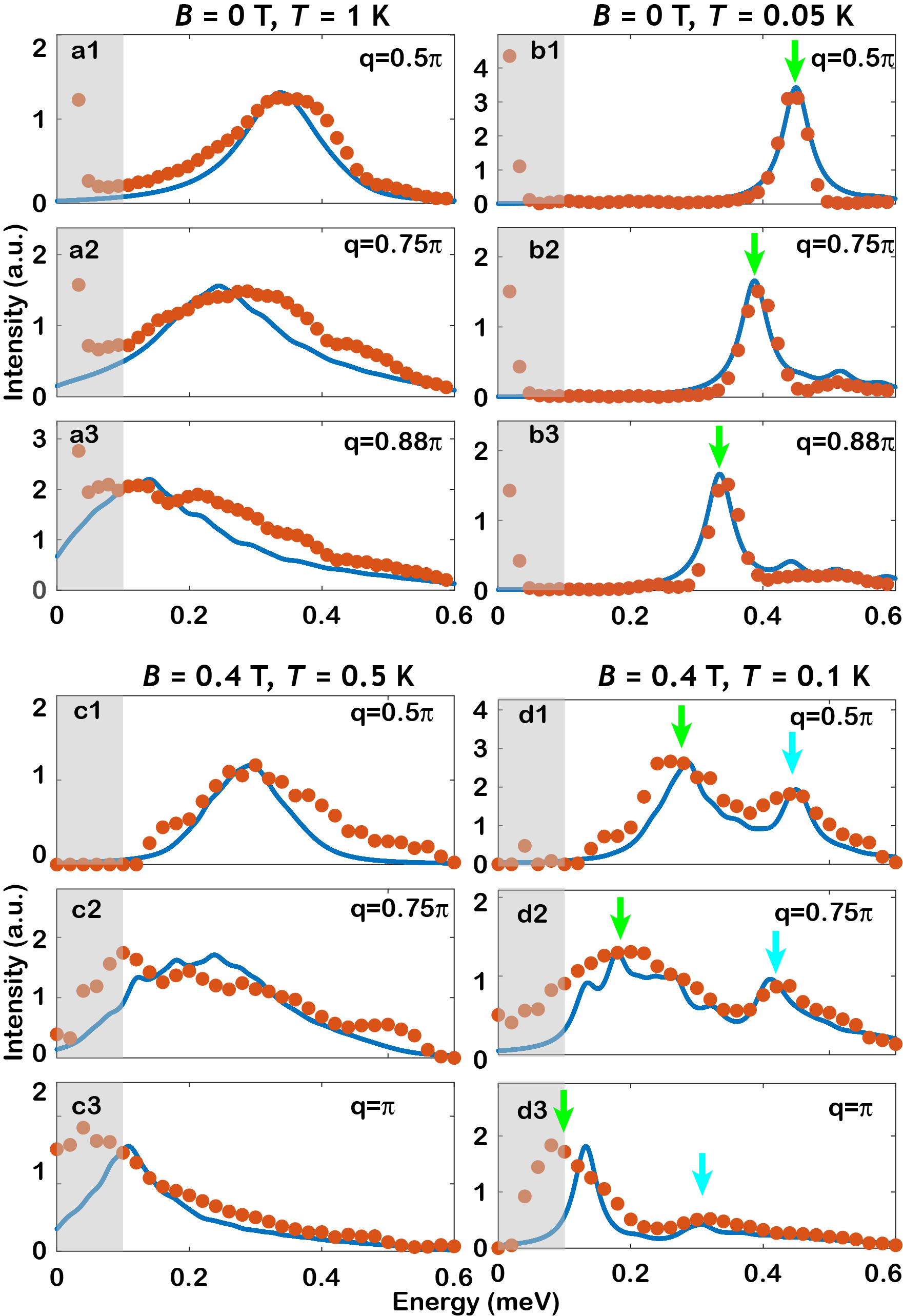}
    \caption{
    Comparison of the INS scattering intensity with theoretical calculations.
    Energy cuts at constant momentum transfers for different magnetic fields and temperatures.
    Red points and blue lines are results from INS measurements and Bethe-ansatz calculations, respectively.
    The green and cyan arrows indicate the positions of confined 1-strings and 2-strings, respectively.
    The INS data in light-gray regions ($\lesssim$ 0.1 meV) are contaminated by elastic line contribution.
    }
    \label{fig:Ecuts_compare}
\end{figure}

The energy cuts at constant momenta resolve the continua and peaks of Bethe strings in Fig.~\ref{fig:Ecuts_compare}.
The broad peaks of the 1-string continuum are well captured by theoretical calculations
with consideration of thermal fluctuation (Fig.~\ref{fig:Ecuts_compare}\textbf{a1-3}).
From the AFM to the SDW phase,
confined 1-strings (Fig.~\ref{fig:Ecuts_compare}\textbf{b1-3})
are suppressed by a magnetic field and give way to
the confined 2-strings at high-energy regions
(Fig.~\ref{fig:Ecuts_compare}\textbf{d1-3}).
While in the disordered region at 0.4 T,
the confinement effect disappears, and
the deconfined 2-strings are hidden inside the 1-string continuum
(Fig.~\ref{fig:Ecuts_compare}\textbf{c1-3}).
Note that the region with energy $\gtrsim$ 0.1 meV,
the theoretical results quantitatively agree with the experimental data.
Especially, the double peaks in experimental data
are evidently confirmed to be the confined 1- and 2-strings
(indicated by green and
cyan arrows in Fig.~\ref{fig:Ecuts_compare}\textbf{d1-3})

{\it Conclusion.~}To conclude, we establish a unified physical picture based on Bethe string
to understand spin dynamics in different phases of YAO.
In the AFM phase, excitations of confined 1-strings are predominant
in the whole dynamical spectrum, while confined 2-strings
take control of the high-energy branch of spin dynamics in the SDW phase.
The excellent comparisons between theoretical and experimental results explicitly
justify the validity of the Bethe-string-based picture
and further confirm
the confinement of many-body Bethe strings in real material.
Following the picture, it allows a uniform characterization of spin dynamics at different phases of a material,
which is beyond the conventional understanding based on static order parameters and low-energy effective theories.
Our study
may also inspire research on non-integrable magnetic systems,
and potentially provides a path toward
a more complete understanding of the many-body
quantum magnetism.

{\it Acknowledgements.~}We acknowledge R. Yu for helpful discussion and R. Jiang for assistance in plotting Fig.1\textbf{a}. This work at Shanghai Jiao Tong University is supported by National Natural Science Foundation of China No. 12274288 and the Innovation Program for Quantum Science and Technology Grant No. 2021ZD0301900 and the Natural Science Foundation of Shanghai with grant No. 20ZR1428400 (J.Y. and J.W.).
Work at Oak Ridge National Laboratory (ORNL) was supported by the U.S. Department of Energy (DOE), Office of Science, Basic Energy Sciences, Materials Science and Engineering Division. We thank J. Keum for assistance with X-ray Laue measurements.
S. E. N. acknowledges financial support from innovation program under Marie Skłodowska-Curie Grant No. 884104.
This research used resources at the Spallation Neutron Source, a DOE Office of Science User Facility operated by the Oak Ridge National Laboratory. X-ray Laue measurements were conducted at the Center for Nanophase Materials Sciences (CNMS) (CNMS2019-R18) at ORNL, which is a DOE Office of Science User Facility.

\appendix

\normalem
\bibliography{bib_confi_string}

\foreach \x in {1,2}
{
    \clearpage
    

\section*{Supplementary material}

# Supplemental Material − Confinement of many-body Bethe strings

Jiahao Yang,[1] Tao Xie,[2] S. E. Nikitin,[3, *] Jianda Wu,[1, 4, 5, †] and A. Podlesnyak[2]

[1]*Tsung-Dao Lee Institute, Shanghai Jiao Tong University, Shanghai 201210, China*
[2]*Neutron Scattering Division, Oak Ridge National Laboratory, Oak Ridge, TN 37831, USA*
[3]*Quantum Criticality and Dynamics Group, Paul Scherrer Institut, CH-5232 Villigen-PSI, Switzerland*
[4]*School of Physics and Astronomy, Shanghai Jiao Tong University, Shanghai 200240, China*
[5]*Shanghai Branch, Hefei National Laboratory, Shanghai 201315, China*

## I. TRUNCATED STRING STATE SPACE METHOD

The one-dimensional spin-1/2 Heisenberg model, the $H_0$ in the main text, can be exactly solved by the Bethe-ansatz method [1]. Following the string hypothesis [2–4], all eigenstates of $H_0$ can be considered as the combination of different numbers of strings with different lengths, as mentioned in the main text. The length of the string can be taken from 1 to $N$ (number of lattice sites). While in most cases, we are concerned with excitations with energies of at most a few $J$, which contain mostly 1-strings and a few non-trivial $n$-strings ($n \geq 2$). Henceforth, we can consider the truncated sub-Hilbert space spanned by those states to study the physics of Hamiltonian like $H_0$ and $H$ in the main text. More details can be found in the reference [5].

## II. ADDITIONAL CALCULATION RESULTS AND COMPARISONS

The zero-temperature dynamical structure factor (DSF) of $H_0$ at zero magnetization is presented in Fig. S1**a**, which is compared with experimental data in Fig. S1**b**. The comparison obviously reflects the essential thermal fluctuation at 1 K.

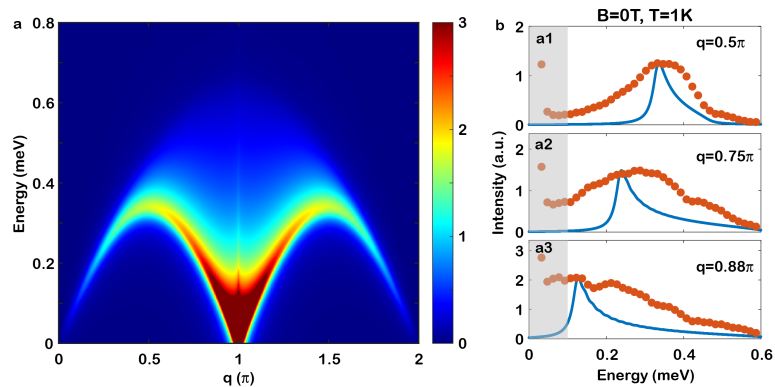

FIG. S1: **a** The zero temperature DSF for the Hamiltonian $H_0$ at zero magnetization. **b** Corresponding comparisons with INS experimental data.

With $m = 12\%$, the zero temperature DSF of $H_0$ and the contributions of it from different Bethe strings are listed in Fig S2.

With $m = 12\%$ and $h_Q = 0.35J$, the zero-temperature DSF of $H$ and the contributions of it from different confined Bethe strings are listed in Fig. S2.

For a better view of satellite peaks, the momentum cut at zero energy is shown in Fig. S4.

---

*Electronic address: stanislav.nikitin@psi.ch
†Electronic address: wujd@sjtu.edu.cn

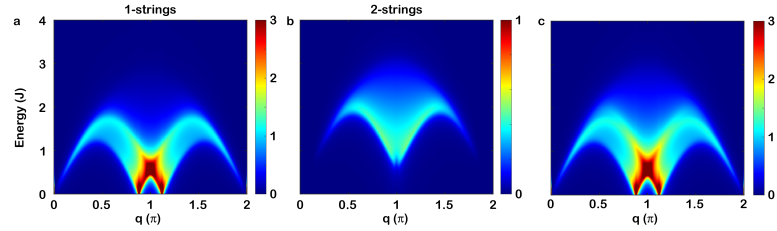

FIG. S2: The zero temperature DSFs of $H_0$ with $m = 12\%$. **c** is the combination of **a** and **b**.

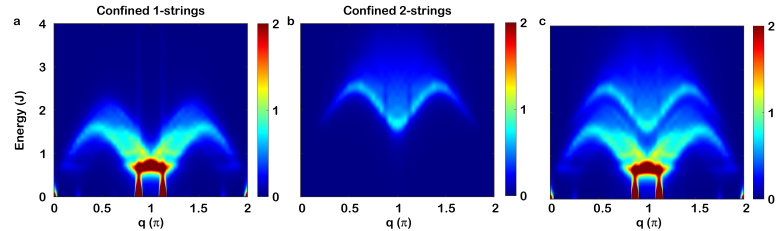

FIG. S3: The zero temperature DSFs of $H$ with $m = 12\%$ and $h_Q = 0.35J$. **c** is the combination of **a** and **b**.

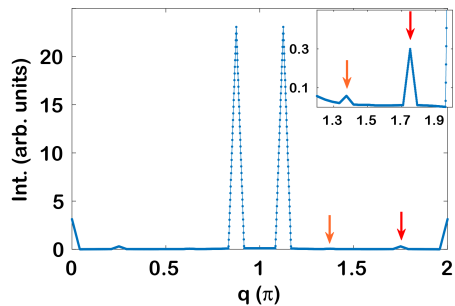

FIG. S4: The satellites at zero energy. The orange and red arrows indicate the position of the emergent satellites due to the ordering field. And the inset is shown for better visibility of them.


}

\end{document}